\documentclass[conference]{IEEEtran}
\IEEEoverridecommandlockouts
\usepackage{cite}
\usepackage{amsmath,amssymb,amsfonts}
\usepackage{algorithmic}
\usepackage{graphicx}
\usepackage{textcomp}

\usepackage{amssymb}
\usepackage{pifont}

\usepackage{xcolor}
\def\BibTeX{{\rm B\kern-.05em{\sc i\kern-.025em b}\kern-.08em
    T\kern-.1667em\lower.7ex\hbox{E}\kern-.125emX}}
\begin{document}

\title{Review of Advanced Monitoring Mechanisms in Peer-to-Peer (P2P) Botnets}

\author{\IEEEauthorblockN{1\textsuperscript{st} Wong Yan Shen}
\IEEEauthorblockA{\textit{National Advanced IPv6 Center} \\
\textit{Universiti Sains Malaysia}\\
Penang, Malaysia\\
yanshen@student.usm.my}
\and
\IEEEauthorblockN{2\textsuperscript{nd} Selvakumar Manickam}
\IEEEauthorblockA{\textit{National Advanced IPv6 Center} \\
\textit{Universiti Sains Malaysia}\\
Penang, Malaysia\\
selva@usm.my}
\and
\IEEEauthorblockN{3\textsuperscript{st} Mahmood A. Al-Shareeda}
\IEEEauthorblockA{\textit{National Advanced IPv6 Center} \\
\textit{Universiti Sains Malaysia}\\
Penang, Malaysia\\
alshareeda022@gmail.com}
}

\maketitle

\begin{abstract}

Internet security is getting less secure because of the
existing of botnet threats. An attack plan can only be planned out to take down the botnet after the monitoring activities to understand the behaviour of a botnet. Nowadays, the architecture of the botnet
is developed using Peer-to-Peer (P2P) connection causing it to be harder to be monitored and track down. This paper is mainly about existing botnet monitoring tools. The purpose of this paper is to study the ways to monitor a botnet and how monitoring mechanism works. The monitoring tools are categorized into active and passive mechanism. A crawler is an active mechanism while sensor and Honeypot are the passive mechanisms. Previous work about each
mechanism is present in this paper as well.

\end{abstract}


\section{Introduction}

A botnet is the network of bots. Bot is the naming for infected computers acting as a controlled robots or machines for criminal, known as bot master to carry out cyber-crime like infecting computers and machines with viruses and malware, accessing a webpage at the same time causing denial of service(DOS), important personal details thieving, spamming and the other cyber-crimes \cite{falliere2011sality,enterprise2018internet,wyke2012zeroaccess}.

A user gets infected when they access to a malicious email attachment, visit a compromised website, or unsuspectingly download the bot files onto their computer. Once infected, the botmaster will gain access to the victim’s computer, without acknowledgement from the victim. By accessing to a victim’s computer will allow the botmaster to obtain two critical resources: CPU power and IP address \cite{goodman2017survey}. Combining all the CPU resources from the bots can form powerful equipment equivalent to a modern-day supercomputer service is provided to botmaster. The diversity of IP address will causes confusion and increases the difficulty of a cyber-crime detection as all those IP addresses seems to be legitimate. According to industry estimates, botnets have caused over \$110 billion in losses globally \cite{goodman2017survey}. The Internet security issues caused by botnet has even triggered the US government and American Industries awareness and listing it as the greatest threats to cyber security \cite{lutscher2020home,marinho2017exploring}. According to the improved speed of the Internet and the technology, the usage of Internet of Things(IoT) gadgets has been targeted as victims to cybercrime too \cite{1,2,3,4,5,6,7,8,9,10,11,12,13,14,15,16,17,18,19,20,21,22,23,24,25}. The situation went more serious when IoT manufacturers often neglect security in the various phases of Software development Lifecycle (SDLC) \cite{gopal2018mitigating}.
In order to take down the botnet, long-term monitoring is required to be done before an attack strategy and method can be determined \cite{stutzbach2008characterizing}. Botnet monitoring can be done through some monitoring mechanisms. These monitoring mechanisms are Crawler, Sensor, and Honeypot. Each of these tools is having different functionality in botnet monitoring activities. Since botnet is an important asset for botmaster, so, botmaster has included some anti-monitoring mechanism in their botnet design to prevent being tracked and removed threats from harming their botnet \cite{andriesse2013highly,karuppayah2015zeus,vasilomanolakis2017trust}. And, these antimonitoring mechanism has increased the level of difficulty
for monitoring tools to continuously monitor on the botnet due to the existence of those interferences and disruption \cite{andriesse2015reliable,bock2018next,karuppayah2016boobytrap}.

Brief description of the botnet monitoring activities as follow. Honeypot is normally using in monitoring activities to obtain the bootstrap nodes (first node list) for a crawler to crawl through the botnet and the entrance point for the sensor to be inserted into. Once all the mechanism is in place, the monitoring activities might take up a few weeks or months. It is important to understand how the botnet works in the design of the architecture and mechanism. Then, an equivalent attack plan can be prepared accordingly. The attack plan aims to bring
down all the nodes in a botnet at once because even with a few nodes leave behind might provide chances for the botnet to recover after sometimes \cite{plohmann2011botnets,karuppayah2015zeus}.

The rest of this paper is structured as follows. Section \ref{sec:back} describes the background of the botnet architecture and mechanism. Section \ref{sec:moni} shows the description of botnet monitoring mechanisms. Section \ref{dis} discusses the advantage and disadvantage of these mechanisms. Lastly, Section \ref{sec:con} concludes this paper and discussion about possible future works.

\section{Background} \label{sec:back}

During forming an attack plan, the network analyst needs to understand botnet behaviour as to design a matching attack plan. Therefore, to understand botnet behaviour, we have to know more about botnet architecture and its mechanism. In this section, the botnet architecture and mechanisms are described. Since P2P architecture is the main trend of botnet architecture design, so the focus will be on P2P botnet architecture and mechanisms only.

\subsection{Botnet Architecture}
There are two types of botnet architecture: centralized and decentralized or known as Peer-to-Peer(P2P). The interaction between the bots or the connection between the bots and the server are vary with different botnet architecture depending on the command-and-control (C2) protocol \cite{wang2008advanced}. In early botnet design, botnet structure is designed using only
centralized architecture. However, due to centralized architecture suffers single point of failure, Peer-to-Peer (P2P) architecture is much preferred in botnet architecture design nowadays \cite{rawat2018survey}.

In P2P architecture, there are structured and unstructured categories. Unlike centralized botnet, in P2P structured botnet, bots are able to communicate to each other using P2P protocol
for the purpose to update its neighbour peer information. Then, the botnet connection is fixed and does not change simplily. Gnutella and Kademlia botnet \cite{holz2008measurements,kang2009towards} are examples of P2P structured botnet. In P2P unstructured botnet, there are no fix formation for this type of botnet. Hence, it is very hard to be monitored. Sality, ZeroAccess, Kelihos are some
examples of P2P unstructured botnet that is still active in action till year 2013 \cite{rossow2013sok}.

\begin{itemize}
    \item P2P Structured Botnet:  Structured P2P botnet is build up by the Distributed Hash Table (DHT) decentralized distributed system that provide lookup service by hash table (key, value) pairs and store to the DHT of every bots. The botnet conenction is formed from the indexes (For example, <Kn,val>). The key(Kn) can be representing the IP address of the destination or the distance of the node. Then, the value(val) can be the command from the botmaster. The best example of this type of botnet is Gnutella and Kademlia botnet \cite{holz2008measurements,kang2009towards}. Since the connection is based on hash table, so the architecture is structured and connection is built up from a basis unlike unstructured P2P botnet which does not have a fix structure.

     \item P2P Unstructured Botnet: Unstructured botnet do not form the botnet structure using DHT system like structured P2P botnet. The connection in the unstructured botnet can be modified by botmaster through Membership Maintenance (MM) mechanism to manipulate a new peer list or a Neighbour List (NL) of the bots as to change the formation of the botnet connection \cite{karuppayah2018advanced,vasilomanolakis2017trust}. All bots NL is handled by Membership Maintenance (MM) mechanism which the purpose is to control the entrance of the nodes within the bot. Thus, the overlay connection is flexible and easily modified. When there are nodes that are inactive, the MM mechanism will replace those inactive nodes with a more active nodes as to maintain the botnet structure.
     \item  Challenges in P2P Botnet Monitoring: There are many challenges in P2P botnet monitoring activities. Antimonitoring mechanism designed in the botnet by botmaster has always causing the most troublesome experience for botnet monitoring activities \cite{andriesse2013highly,karuppayah2015zeus}. Anti-monitoring mechanism purpose is to prevent monitoring mechanism like sensor and crawler to be inserted into botnet networks. Anti-monitoring mechanism like IP Filtering and NL controlling are mechanisms that specifically designed to block the monitoring mechanism from entering the botnet and causing interferences to prevent nodes full NL information from flowing out easily. Hence, it is important to know and upgrade the botnet monitoring mechanism to against the anti-monitoring mechanism.
\end{itemize}

\section{Botnet Monitoring Mechanisms}\label{sec:moni}

There are three types of botnet monitoring tools: Honeypot,
Crawler and Sensor. Honeypot and Sensor are passive mechanism as these tools always wait for connection from the bot. But, for active mechanism - Crawler, it will act actively by probing the available nodes for the Neighbour List (NL) \cite{karuppayah2016advanced}. So that, from the list, it can identify the next node to probe.

\subsection{Honeypot}
Honeypot is a tool that used for trapping bots from botnet during attacks to extract information, activities and analyses them in order to understand the botnets \cite{akkaya2012honeypots}. It usually
represent itself in the way of server or high value asset but vulnerable in the system designed. This is to make it to become a potential target for network attackers and then gather information for network security researchers. From the information, researchers can understand more about the techniques and mechanism used in the botnet but at the same
time, they need to prevent botnet from gaining access to the main systems \cite{lee2020review}.

There are 3 types of Honeypot level of interactions: low interaction honeypot, medium-interaction honeypot and high interaction honeypot. The categories are defined based on the services or interaction level provided by Honeypot to potential hackers.

\begin{itemize}
    \item Low Interaction Honeypot: The risk of this Honeypot causing the whole server to compromise to the botnet is low due to it is just a small pile from the server and there is no operating system for the botnet to deal with \cite{lee2021classification,peter2011practical}. Some examples of the low interactive Honeypot are Honeyd, HoneyRJ \cite{peter2011practical}, BotMiner, BotGrep and BotTrack \cite{haltacs2014automated}.

    \item Medium Interaction Honeypot: Medium interactive Honeypot is better in terms of the information that it can obtain from the mechanism. That is because it is designed to have more security holes so that bots from botnet can access to the system \cite{peter2011practical}. Hence, more information and complicated attacks from the botnet can be gathered compare to low interactive Honeypot. Mwcollect, Honeytrap, Nepenthes \cite{akkaya2012honeypots}, HoneyBOT \cite{peter2011practical}  and Kippo Honeypot Distro \cite{dowling2017zigbee} are some of the medium interaction honeypots that are used today.

    \item High Interaction Honeypot: Unlike the other two, there is an operating system in this type of Honeypot. Hence, the botnet can perform any activities in this mechanism. Thus, more data can be gathered from the botnet activities through high interactive Honeypot. However, it is the most risky one when it comes to security as it provides such loosen access to the botnet and there is little to no restrictions placed on what the bots can do once compromises the system. 
    
    The most significant weakness of this Honeypot is time consuming and difficult to maintain \cite{akkaya2012honeypots,dowling2017zigbee,karuppayah2018advanced,peter2011practical}. Specter and Argos are good examples of a high interaction honeypot \cite{dowling2017zigbee,peter2011practical}. Besides that, there are Minos, ManTrap and Argos \cite{akkaya2012honeypots}.
\end{itemize}

\subsection{Crawler}

Generally, crawler is acting as an active mechanism in the botnet monitoring tools. It starts by requesting NL from the bootstrap nodes by probing on those nodes in active. Once the NL is received, it will continuously requesting for more nodes
from all the active nodes in the NL until all nodes are being discovered or the action is being terminated \cite{karuppayah2018advanced}. The crawling method is mostly developed using Breadth-First-Search (BFS) or Depth-First-Search (DFS) method \cite{spampinato2019linear}. Throughout the process, it would be able to reveal the identity of participating bots in the botnet. With all the information available, the botnet size can now be estimated and enumerated. This is because crawler is using graph traversal techniques to request nodes for NL connectivity. Hence, this allows analyst to reconstruction of the connectivity graph and the topology of the botnet \cite{haas2016resilience,karuppayah2014advanced,karuppayah2016boobytrap}. The following subsection will be some description on different crawlers designed by other.

\begin{itemize}
    \item P2P Graph Search Method: Graph search crawling
method target is to reconstruct the P2P botnet graph by visiting
all the nodes and requesting their peer lists. Then, further
action is done through graph search to attack against the P2P
botnet \cite{rossow2013sok}. It is initialized with a set of seed peers, which obtained through reverse engineering bot samples or dynamic analysis. P2P botnet topologies are so dynamic that it keeps
on changing and affecting the output results of the crawler.
Furthermore, during crawling, it relies on the seed peer list,
some peers may not be able to be discovered via graph search
method because there might be churn \cite{plohmann2011botnets} or peers behind
Firewall and Network Address Translation (NAT) devices.

    \item Storm Crawler: The crawler design in \cite{holz2008measurements} purpose is
to understand more on the peer’s connection in the botnet.
The crawler design is based on BFS method to locate all the
involving peers. In order to avoid overlapping results from
the peer’s response, the system will send 16 route requests
to every peer, and categories them as different zones of the
peer’s routing tree. For each of the node that have response,
the system will add to the peer list of its routing tree. But the
next node to crawl on will according to the sequence according
to First In First Out (FIFO) concept.

    \item  Nugache Crawler: In \cite{dittrich2008discovery} paper, the author has designed the crawler as an enumeration tool to estimate the botnet Nugache size while avoid getting notice from the botnet operator. By using the DFS method, the crawler will keep establishing with the next level nodes unless there is no more nodes in the branch. DFS crawling method will be looking into deep down starting with first node follow by the node in the next level, unlike BFS which go for the next side node. DFS will always complete the routing connection of the first node then only it will move on to the side node. The crawling algorithm is based on Last-In-First-Out (LIFO) concept.

\item  Less Invasive Crawling Algorithm(LICA) crawler \cite{karuppayah2014advanced}:
LICA crawler design is focusing on crawling efficiency and
claim to be able to adapt to different environment by using
the parameters calculation. The crawler will crawl from node
in bootstrap list and limit the crawl number up to a parameter.
The crawling activities ended when all contactable nodes in
the network have been discovered or the limit set has reached.
If the limit has not reached, LICA crawler may repeat another
iteration of crawling. However, it will use the information
from previously segregated nodes instead of the “seedpeer”
list as the crawling list. Overall, the concept of LICA
crawler is to crawl over those popular nodes and ignoring
those less connected nodes as to cover the botnet connection,
as larger as possible for monitoring the botnet but not in detail.

\end{itemize}

\subsection{Sensor}

Sensor node is acting passively after deployment in the
botnet, waiting for connection from other bots. If sensor
node keeps presenting and responding to other bots, it will
be upgraded to a popular superpeer and will be able to mix
in to the bots prefered member list. The bots will start to
share around about this sensor node \cite{karuppayah2017sensorbuster}. Then, it will be
able to observe the messages flowing around the botnets
about the active nodes in the botnet. From these messages,
network security analyst will know which of the nodes that
are involved in the botnets \cite{haas2016resilience,karuppayah2018advanced}. This is the main purpose of a sensor which is to be as popular as possible so that to be able to increase their visibility as higher as possible. Although
sensors able to obtain information about the bots, they cannot
gather finegrained data or to obtain the interconnectivity of
the bots \cite{haas2016resilience}. The following subsection will be describing some botnet monitoring researches involving sensor node.

\begin{itemize}
    \item Monitoring Kelihos Botnet: Starting from the bootstrap
nodes in Kelihos botnet domains, they can track the growing
population of hosting IPs and detect new fast flux domains
hosted by Kelihos botnet if the new domains match with
the monitored DNS authoritative traffic. This will allows the
author to analyse on various components and attributes of the
infrastructure used by the Kelihos fast flux botnet. During the
monitoring, they also included filtering function in order to
avoid false positives result and confused by other non-botnet
fast flux domains \cite{marinho2017exploring}.

    \item Monitoring on Peer-to-Peer(P2P) Botnet: Based on
Rossow et al. \cite{rossow2013sok}  observation, P2P botnets peers are periodically contacted by the neighbouring peers especially during
regular peer list verification cycles. Besides that, sensor can
also be contacted by non-routable peers unlike crawler, which
then make sensor able to enumerate more nodes in a botnet. From the research results, sensor nodes have revealed large
number of bots which cannot be found by crawlers. Another
methodological difference is that, crawlers are actively enumerating
peers, while sensors are working passively that it
waits to be contacted by bots. And, sensor’s coverage relies
on its popularity in the botnet. So, the sensor node requires
more time to leverage the outcome of the monitoring result.

    \item Passive P2P Monitoring in Storm Botnet: Passive P2P
Monitoring (PPM) is a botnet monitoring tool that designed
by Kang et al. \cite{kang2009towards} that will act as peer nodes, primarily in Storm botnet’s P2P network. The sensor node will only listens in the Storm botnet and pretending itself as a legitimate bot
and routes messages. PPM might only have a small part of information about the whole network, it knows only those nodes in its routing table and those nodes that contacted it. This is because it only listens and will never contact with other nodes by itself due to the passive nature. PPM can identify a node location regardless it is behind a NAT or a Firewall. Then, it will further distinguishes the node into different class. This is a function which sensor node overtake crawler functionality \cite{kang2009towards}.
   \end{itemize}

\section{Advantage and Disadvantage }\label{dis}

Table \ref{tab:honeybot} shows the comparison between different Honeypots according to their advantages and disadvantages described in
different researches. Nowadays Honeypots has utilized faster networking and visualization technologies during developing Honeypot. And, they have also deployed high interaction Honeypots and
low risk Honeynets, isolating attack traffic from connected
hardware and networks. It is good to see a lot of
improvements that have been done in this mechanism as to
stop and prevent network security issues.

\begin{table}[h]
	\caption{Comparison Between Different Honeypots}
	\centering
	\renewcommand{\arraystretch}{1.3}
 	\begin{tabular}{p{25pt}p{90pt}p{90pt}}
		\hline
	Honeypot Types 	&Advantages & Disadvantages \\
		\hline
Honeyd (Low Interactive) & \begin{itemize}
    \item Able to act like normal operating system.
    \item Able to create and run with fake IP addresses simultaneously.
\end{itemize} & \begin{itemize}
    \item Aged and outdated NMAP fingerprint which will expose the identity of the honeypot.
    \item The operating scripts are bound to different ports.
\end{itemize} \\
HoneyRJ (Low Interactive)&  \begin{itemize}
    \item Extremely simple and easily extendable.
    \item Multi-threaded to support multiple connections.
\end{itemize} & \begin{itemize}
    \item only supports string-based protocols and does not support the transmission of binary data.
  
\end{itemize} \\
Kippo (Medium Interactive) &\begin{itemize}
    \item Able to simulate other file systems and log all interactions with automated or non-automated attacks.
    \item Utilized IoT device’s vulnerability as lure to hacker.
\end{itemize} & \begin{itemize}
    \item only provide partial implementation of services and do not allow full interaction with the system.
   
\end{itemize}\\
Nepenthes ( Medium  Interactive )& \begin{itemize}
    \item Having Scalability and flexibility features.
    \item Able to create many honeypots in the system and capture data easily.
\end{itemize}& \begin{itemize}
    \item Many restrictions on the setting.
\end{itemize}\\
Honeywall ( High Interactive) &\begin{itemize}
    \item Hard to be hacked by hacker into Honeywall
administration system.
    \item Able to capture more useful and interesting findings.
\end{itemize} & \begin{itemize}
    \item Implementation will be time consuming and complicated.
    \item Does not hide the network address.
\end{itemize}\\
		\hline
		
	\end{tabular}
	\label{tab:honeybot}
\end{table}

Table \ref{tab:crawlers} shows the comparison between different types of crawler about their strength and weakness. Most of the crawler
seems lacking of features to avoid anti-monitoring mechanism
and churn effects. But, it might be due to some of the crawlers
are designed long before the existence of those interferences. Crawlers in future might need to include features that
can tolerate with anti-monitoring mechanism affects and have
churn consideration like crawler mentioned in \cite{rossow2013sok}. These
features are important for crawler to be efficient because
botnet nowadays is having those anti-monitoring mechanism
and performance of crawler without those features might get
affected and the monitoring results might be degraded as
mentioned in \cite{andriesse2015reliable,karuppayah2016boobytrap,vasilomanolakis2017trust}.

\begin{table}[h]
	\caption{Comparison Between Different Crawlers}
	\centering
	\renewcommand{\arraystretch}{1.3}
 	\begin{tabular}{p{25pt}p{90pt}p{90pt}}
		\hline
	Crawler Types 	&Advantages & Disadvantages \\
\hline
P2P 		& Covered for all P2P botnet evaluation and operated in real-time& Not capable to handle anti-monitoring\\
Storm 		& BFS crawler, easy to design and crawl fast& Not capable to handle anti-monitoring mechanisms and churn effects\\
Nugache 		& DFS crawler, easy to design&Not capable to handle anti-monitoring mechanisms and churn effects \\
LICA 		& Fast crawling algorithm
different from BFS and DFS& Easily missed out nodes during churn effects\\
\hline
	\end{tabular}
	\label{tab:crawlers}
\end{table}

Table \ref{tab:sesnro} shows the comparison between different sensor in different botnet monitoring. Through the experiment result in \cite{kang2009towards,rossow2013sok}, it shows that more than 40\% of bots that contact the sensor are behind firewall or NAT devices. Since crawler is unable to reach those nodes, it is wise to combine crawlers and sensor nodes in a monitoring activities as to provide much more accurate population estimation by using advantage of a sensor to complement the weakness of crawler \cite{rossow2013sok}. It will helps network analyst to obtain a better picture of a botnet connection.

\begin{table}[h]
	\caption{Comparison Between Different Sensors}
	\centering
	\renewcommand{\arraystretch}{1.3}
 	\begin{tabular}{p{25pt}p{90pt}p{90pt}}
		\hline
	Sensor Types 	&Advantages & Disadvantages \\
\hline
Kelihos Botnet & Able to monitor on fast flux spam domains and process on it. Able to eliminate the false positive result by itself.& Not considering anti-monitoring mechanisms and churn effects during the monitoring activities\\
PPM in
Storm Botnet&Able to trace nodes behind NAT devices and Firewall & Could not find nodes with short life-time\\
Monitoring
P2P Botnet & Able to trace nodes behind NAT devices and Firewall even in most of the P2P botnet & Sensor’s coverage depends on its popularity \\

\hline
	\end{tabular}
	\label{tab:sesnro}
\end{table}

\section{Conclusion} \label{sec:con}

Although it seems like the three monitoring tools are
sufficient for botnet monitoring purpose. However, there are
some weakness in each of the monitoring tools. For Honeypot,
cost of maintaining a effective honeypot can be high because
the specialized skills are required to implement it. Or else,
it might causes vulnerability in the system configuration and
lead to breach in the system.

The P2P botnet crawling mechanism will only able to crawl
on the nodes before NAT devices and Firewall \cite{plohmann2011botnets}. That is due to the mechanism of NAT devices and Firewall will filtering out the unknown address that trying to pass through them and block out the crawler. The disadvantage of sensor node would be its passive behaviour. As it might take long time to have some results because it need to become popular and always online to be able to contribute to the P2P network \cite{karuppayah2018advanced}. Beside that, sensor is unable to determine if the node that approach it is using spoofed or real IP address \cite{kang2009towards}. Hence, it is advisable to combine crawler and sensor.

Furthermore, with the improvement of the botnet
antimonitoring mechanism mentioned in works done in \cite{rossow2013sok,karuppayah2017sensorbuster,vasilomanolakis2017trust}, and it is worst when botmaster can even backtrace
the monitoring mechanisms existence in the botnet \cite{karuppayah2016boobytrap}. Then,
botmaster might even launch a retaliation attack on the
monitoring system for revenge \cite{andriesse2013highly}. There might be one day all these monitoring tools will be blocked by the botnet.
Nonetheless, there is forecast that the architecture of advance
botnet like Zeus and Sality are highly resilient to sinkholing
attacks \cite{rossow2013sok}. Hence, there might be potential threats from P2P
botnet where all the existing methods might not be able to
monitor or take down the botnet in future, if all these tools and
methods cannot be improved or alternate mitigation methods
cannot be discovered soon enough.

	\bibliographystyle{unsrt}
	\bibliography{ref}
\end{document}